\documentclass[11pt]{amsart}
\usepackage{amscd,amssymb}
\usepackage{amsmath}
\usepackage{amsthm}
 
\usepackage[scaled]{helvet} 
\usepackage[pdftex]{graphicx,color}

\usepackage[cmtip,matrix,arrow]{xy}
\numberwithin{equation}{section}

\newtheorem {thm} 			{Theorem}     

\newtheorem {thm_}                      {Theorem}     
\newtheorem {cor_}	[thm_]	        {Corollary}
\newtheorem {conj_}     [thm_]       	{Conjecture}

\newtheorem {thm_section} 		{Theorem} [section] 
\newtheorem {prop_section}[thm_section]	{Proposition}
\newtheorem {cor_section} [thm_section]	{Corollary}

\theoremstyle{definition}

\newtheorem {rem}       [equation]	{Remark}

\newtheorem {ex}        [thm_section]	{Example}

\newcommand{\pr} {\smallskip\noindent{\bf Proof\,\,}}
\newenvironment{Proof}  {\pr}{\hspace*{\fill}\qed\\}

\DeclareMathOperator{\Thet}{\Theta}
\DeclareMathOperator{\C}{C}
 \DeclareMathOperator{\I}{I}
 \DeclareMathOperator{\J}{J}
\DeclareMathOperator{\Ham}{H}
\DeclareMathOperator{\Z}{Z}
\DeclareMathOperator{\G}{\Gamma}
\DeclareMathOperator{\Ps}{\Psi}
\DeclareMathOperator{\Fi}{\Phi}
\DeclareMathOperator{\thet}{\theta}
\DeclareMathOperator{\Sym}{Sym}
\DeclareMathOperator{\Rep}{Re}

\DeclareMathOperator{\supp}{supp}
\DeclareMathOperator{\trace}{tr}

\newcommand{\Str}{S^1}

\newcommand{\Cont}{\mathbf{C}}
\newcommand{\Lp}  {\mathbf{L}}
\newcommand{\Hsp} {\mathbf{H}}

\newcommand{\Intg}{\mathbb{Z}}
\newcommand{\Real}{\mathbb{R}}
\newcommand{\Cmpl}{\mathbb{C}}

\newcommand{\Sch} {\mathcal{S}}
\newcommand{\Ban} {\mathcal{B}}
\newcommand{\Hilb}{\mathcal{H}}
\newcommand{\Fock}{\mathcal{F}}
\newcommand{\El}{\mathcal{L}}

\newcommand{\bra}{\Bigl \langle}
\newcommand{\ket}{\Bigr \rangle}

\newcommand{\ip}{\bra \, , \, \ket}

\providecommand{\abs}[1]{\lvert#1\rvert}
\providecommand{\norm}[1]{\lVert#1\rVert}

\begin{document}

\title[Nonperturbative String Field Theory]{Measures on Banach Manifolds, Random Surfaces, and Nonperturbative String Field Theory with Cut-offs}


\author{Jonathan Weitsman}
\address{Department of Mathematics, University of California, Santa Cruz, CA 95064}
\curraddr{}
\email{weitsman@math.ucsc.edu}
\thanks{Supported in part by NSF grant DMS 04/05670}
\thanks\today
\subjclass[2000]{Primary }


\date{}

\begin{abstract}
We construct a cut-off version of nonpertubative closed Bosonic string field theory in the light-cone gauge with imaginary string coupling constant.  We show that the partition function is a continuous function of the string coupling constant, and conjecture a relation between the formal power series expansion of this partition function and Riemann Surfaces.
\end{abstract}

\maketitle
\section{Introduction}\label{sec:introduction}

Recent work of Schnabl \cite{Schnabl}, confirming conjectures of Sen \cite{Sen},
has given strong evidence for string field theory as a candidate for nonperturbative string theory. This may indicate that string field theory provides a framework for the advances in nonperturbative string theory which have influenced both mathematics and physics.
In view of this fact, and of the important role of nonperturbative string theory in mathematics, it seems reasonable to ask to what extent string field theory can be understood mathematically.
In this paper we will show that in the simple case of the light cone closed Bosonic string field theory, and once appropriate cut-offs have been introduced, such a mathematical interpretation is possible.\footnote{For another mathematical study of string field theory, see Dimock \cite{dimock}.}

\subsection{Free string field theory for the closed Bosonic string:  Formal path integrals}\label{subsec:free-string-field}

Free field theory for the closed Bosonic string in light cone gauge was described by Kaku and Kikkawa \cite{KK}. The string field $\Ps(t,\ell,\phi)$ is formally a complex-valued function of time $t\in \Real$, string length\footnote{In the light cone gauge this string length appears as the momentum conjugate to the light cone coordinate $X^+;$ see \cite{KK}.} $\ell \in (0,\infty)$, and a map  $\phi:\Str \to \Real^d$, which may be thought of as a loop or ``string'' in $\Real^d$ (more properly $\phi$ should be considered as a distribution on $S^1$ with values in $\Real^d$).\footnote{This model corresponds to a gauge-fixed theory of strings propagating in $\Real^{d+1,1};$ see e.g. \cite{polchinski}.} In view of the Schrodinger representation\footnote{In this paper all vector spaces are complex unless specifically denoted otherwise.}
\[
\Lp_2(\Sch'(\Str,\Real)\otimes \Real^d, d\mu_{\sqrt{-(d^2/dx^2) + m^2}} ) \simeq \Fock,
\] 
(where $m > 0$; see Appendix \ref{app:a}) which identifies functions on distributions on $\Str$ with elements of Fock space $\Fock,$ we may formally think of the string field as a distribution $\Psi(t,\ell)$ on $\Real \times [0,\infty)$ with values in some completion of Fock space.

In these terms a cut-off version of Kaku and Kikkawa's \cite{KK} formal free string path integral is given by 
\begin{equation}\label{eq:i1}
\int d\Psi \, e^{  \int_{L_0}^{L_\infty} d\ell \int_{-\infty} ^{\infty} dt \,\bra \Ps(t,\ell), \bigl( i\frac{d}{dt} - \Ham_{\ell,m} \bigr) \Ps(t,\ell) \ket_\Fock  },
\end{equation}
where ${\Ham_{\ell,m}}$ is the Hamiltonian on Fock space of mass $m > 0$ (see Appendix A) and $L_0, L_\infty,$ with $ 0 < L_0 < L_{\infty} < \infty,$ correspond to the maximum and minimum allowed string lengths.

To try to interpret an expression of the type \eqref{eq:i1} as a measure, we compute correlation functions. If we take $t,t' \in ( -\infty, \infty)$, $g,g' \in \Cont^{\infty}([L_0,L_{\infty}])$, and $v, v' \in \Fock$, and set
\[
\Fi_{v,g,t}(\Ps) = \int _ {L_0}^{L_{\infty}} g(\ell) \bra \Ps(t,\ell),v \ket_\Fock \, d\ell,
\]
then the expression \eqref{eq:i1} gives rise to the formal computation
\begin{multline}\label{eq:i2}
\int \overline{\Fi}_{v,g,t}(\Ps) \, \Fi_{v',g',t'}(\Ps) \, d\Psi \, e^{  \int_{L_0}^{L_\infty} d\ell\int_{-\infty} ^{\infty} dt\, \bra \Ps(t,\ell), \bigl( i\frac{d}{dt} - \Ham_{\ell,m} \bigr) \Ps(t,\ell) \ket_\Fock  } \\
=  \int_{\
L_0}^{L_\infty}  \bra v,  e^{ -i(t - t') \Ham_{\ell,m} } v' \ket_\Fock  \thet(t-t') \overline{ g}(\ell) g'(\ell)\, d\ell,
\end{multline}
where the function $\thet: \Real \to \Real$ is defined by
\[
\begin{cases}
\thet(x) =0 \quad \text{if} \quad x < 0;\\
\thet(x) =1 \quad \text{if} \quad x \geq 0.
\end{cases}
\]
\subsection{Free string field theory for the closed Bosonic string:  Mathematical results} 
In Section 2 we will prove the following result, which may be viewed as a mathematical version of \eqref{eq:i2}.  Choose  $\alpha > 0$ and let $\Fock^\alpha = e^{-\alpha \Ham_{L_0,m}} \Fock.$ 
\begin{thm_}\label{thm:i1}
There exists a Banach space $\Ban \supset \Hsp_1(\Real) \otimes \Lp_2([L_0,L_{\infty}]) \otimes \Fock^\alpha$  and a Gaussian probability measure $\mu$ on $\Ban$ such that if $v,v' \in \Fock$ are eigenvectors of $\Ham_{\ell,m}$, $g,g' \in \Cont^{\infty}([L_0,L_{\infty}])$, and $t,t' \in \Real$, the function
$\Fi_{v,g,t}: \Hsp_1(\Real) \otimes \Lp_2([L_0,L_{\infty}]) \otimes \Fock \to \Cmpl$, given by
\[
\Fi_{v,g,t}(f\otimes h \otimes w) = f(t) \int_{L_0}^{L_{\infty}} d\ell\, \overline{h}(\ell)g(\ell)\, \bra w,v \ket_\Fock
\]
extends to an element $\Fi_{v,g,t} \in \Lp_p(\Ban,d\mu)$ for all $p \geq 1$. Furthermore, 
\begin{equation}\label{eq:i3}
\int_{\Ban} d\mu \,  \overline{\Fi}_{v,g,t} \, {\Fi}_{v',g',t'}  = \int_{L_0} ^ {L_{\infty}} d\ell\, \overline{g}(\ell)\, g'(\ell) \, \bra v, e ^ {-\vert t -t'\rvert \Ham_{\ell,m} } v' \ket_\Fock.
\end{equation}
\end{thm_}
\begin{rem} Note that \eqref{eq:i3} differs from \eqref{eq:i2} by the absence of the function $\thet(t-t')$ and by the familiar ``Wick rotation'' $i(t-t') \to (t-t')$. In physical language we have replaced the first order Minkowski action by a second order Euclidean action.  This would seem to be necessary to get a reasonable field theory limit; furthermore the first order theory of Kaku and Kikkawa \cite{KK} does not produce any vacuum-to-vacuum diagrams in either the free or interacting theories.\end{rem}

Our main technique for producing the measure $\mu$ is the abstract Wiener space construction of L. Gross \cite{Gross} (see Appendix \ref{app:b}).

The path integral description of quantum field theory relates the measure $\mu$ to quantities associated with two dimensional quantum field theory on the Riemann surface $S^1 \times \Real$ as follows.  Let $\Str_{\ell}$ denote the circle of length $\ell$, and let $d\nu_{\ell,C}$ be the Gaussian measure on $\Sch'(\Str_{\ell} \times \Real,\Real)\otimes \Real^d$ with covariance $C = (-\Delta + m^2)^{-1}$. Then if $g,g' \in \Cont^{\infty}([L_0,L_{\infty}])$, $t,t'\in \Real$, and $f,f' \in \Cont^{\infty} (\Str,\Real)\otimes \Real^d$, we may define for $\psi \in \Cont^{\infty}(\Str_{\ell} \times \Real,\Real)\otimes \Real^d$
\[
\Fi_{f,g,t}(\psi) = \int_{L_0}^{L_\infty}d\ell\, g(\ell) \int_{\Str_\ell} \bra \psi(s,t) , \frac{1}{\sqrt{\ell}} f(s/\ell) \ket_{\Real^d} ds
\]

Then the map $\Fi_{f,g,t}:\Cont^{\infty}(\Str_{\ell} \times \Real,\Real)\otimes \Real^d\times [L_0,L_\infty] \to \Cmpl$ extends to a function
\[
\Fi_{f,g,t} \in \Lp_p(d\nu_{\ell,C} \times dm_{[L_0,L_\infty]}) {\rm ~for~ all~} p \geq 1,
\]
where $m_{[L_0,L_\infty]}$ is Lebesgue measure on $[L_0,L_\infty].$   And we have 
\begin{equation}\label{eq:i4}
\int_{L_0}^{L_\infty} d\ell \int_{\Sch'(\Str_{\ell} \times \Real,\Real)\otimes \Real^d }\overline{\Fi}_{f,g,t} \, {\Fi}_{f',g',t'} \, d\nu_{\ell,C} = \frac{1}{2}\int_{L_0} ^ {L_{\infty}} d\ell\, \overline{g}(\ell)\, g'(\ell)\, \bra f, \frac{e ^ {-\lvert t -t' \rvert \sqrt{-\frac{1}{\ell^2}(d^2/dx^2) + m^2 }}} {\sqrt{-\frac{1}{\ell^2}(d^2/dx^2) + m^2 } } f' \ket_{L_2(S^1,\Real)\otimes \Real^d}.
\end{equation}
in line with \eqref{eq:i3}.

\subsection{Interacting String field theory:  Formal path integrals}\label{subsec:inter-string-field}
Interacting string field theory is obtained from a cubic function on the space $\Ban$. We first describe a version of the formal construction of \cite{KK}. We imagine a string -- that is, a function $\phi: \Str_{\ell} \to \Real^d$ -- breaking up into two loops 
\begin{align*}
\phi_1&: \Str_{\ell_1} \to \Real^d\\
\phi_2&: \Str_{\ell_2} \to \Real^d
\end{align*} 
where $\ell_1 + \ell_2 = \ell$. In mathematical terms, we have a projection 
\begin{equation}\label{eq:i5}
\pi_{\ell}^{\ell_1,\ell_2} : \Lp_2(\Str,\Real)\otimes \Real^d \to (\Lp_2(\Str,\Real) \otimes \Real^d)\oplus (\Lp_2(\Str,\Real)\otimes \Real^d),
\end{equation}
giving rise to a map\footnote{Note that since Fock space is constructed from the symmetric product of square-integrable loops, which are not continuous, the problem of ``gluing'' two loops of lengths $\ell_1$ and $\ell_2$ to form a loop of length $\ell_1 + \ell_2,$ which is a cause of concern in the physics literature, does not arise.}

\begin{equation}\label{eq:i6}
\pi_{\ell}^{\ell_1,\ell_2} : \Fock \to \Fock \otimes \Fock\end{equation}
and, if we imagine for a moment that the string field $\Psi$ is a {\em function} with values in Fock space, we formally obtain a cubic interaction term
\begin{equation}\label{eq:i7}
\I(\Ps) = \negmedspace \int_{-\infty}^{\infty}\! dt \negmedspace \int_{L_0}^{L_\infty} 
             \negmedspace \int_{L_0}^{L_\infty} d\ell_1 \, d\ell_2
                 \bra 
                   \Ps(\ell_1,t) \otimes \Ps(\ell_2,t) ,
                   \pi_{\ell}^{\ell_1,\ell_2} \Ps(\ell_1 + \ell_2,t) 
                 \ket_\Fock.
\end{equation}

A version of the partition function of the interacting string field theory of \cite{KK} is given formally by
\begin{equation}\label{eq:i8}
\Z(\lambda) = \int d\Psi \, e^{  \int_{L_0}^{L_\infty} d\ell \int_{-\infty} ^{\infty} dt \,\bra \Ps(t,\ell), \bigl( i\frac{d}{dt} - \Ham_{\ell,m} \bigr) \Ps(t,\ell) \ket_\Fock  + \lambda \Rep \I(\Psi)}.
\end{equation}

The expression \eqref{eq:i8} gives rise to a formal power series, each term of which corresponds to a directed trivalent ribbon graph\footnote{Recall that a {\em ribbon graph} is a graph along with an assignment to each vertex of a cyclic ordering of the edges abutting that vertex.} $\Gamma$ whose edges are labeled by two variables $t_{e} \in \Real$, $\ell_e \in [L_0, L_\infty]$. Formally, then,
\[\label{fpe}
\Z(\lambda) \sim 
      \sum \limits_{\G} \frac{|{\rm Aut}(\G)|}{|{\rm vert~}(\G)|!}\lambda^{|{\rm vert~}(\G)|} 
        \int_{-\infty}^{\infty}      \ldots    \int_{-\infty}^{\infty} 
          \prod_{e \in e(\G)} \, dt_{e} 
        \int_{L_0}^{L_{\infty}}      \ldots    \int_{L_0}^{L_\infty}
          \prod_{e \in e(\G)} \, d\ell_{e}
            f_\Gamma(\{t_e\},\{\ell_e\}),
\]  where $|{\rm vert~}(G)|$ is the number of vertices in $\G$ and $|{\rm Aut}(\G)|$ is the order of the group of automorphisms of $\G.$
Kaku and Kikkawa's \cite{KK} arguments lead one to expect that
\[
f_\Gamma(\{t_e\},\{\ell_e\}) \sim \bigl(\det (-\Delta_{\G,\{t_e\},\{\ell_e\}} + m^2)\bigr)^{-d/2},
\]
where $\Delta_{\G,\{t_e\},\{\ell_e\}}$ is the Laplacian on a Riemann surface formed by replacing each edge $e$ of the ribbon graph $\G$ by a tube of length $t_{e}$ and width $\ell_{e}$, and gluing the corresponding tubes to form a two-manifold with conical singularities, and the determinant of the operator $-\Delta_{\G,\{t_e\},\{\ell_e\}} + m^2$ is defined in some appropriate way.\footnote{In Section \ref{rsurf} we make a precise conjecture relating string field theory to the construction by \cite{kokkor} of the determinant of the Laplacian on surfaces with conical singularities.}
\subsection{Interacting String field theory: Mathematical results}
We now turn to a mathematical construction of a cut-off version of the function $\Z(\lambda)$. Let $M >0$ and let $\Fock_M$ denote the finite-dimensional subspace of Fock space $\Fock$ where the Hamiltonian $\Ham_{L_0,m}$ is bounded above by $M$. Let $p_{M}: \Fock \to \Fock$ denote the corresponding projection. Let $\chi \in \Cont_c^{\infty}(\Real)$ satisfy
\begin{itemize}
\item $\chi(-x) = \chi(x)$;
\item $\chi \geq 0$;
\item $\supp \chi \subset [-1,1]$;
\item $\int_{-1}^{1} \chi (x) \, dx = 1$.
\end{itemize}
For $\kappa > 0$ let $\delta_{\kappa}(x) = \kappa \chi (\kappa x)$. 

Let $\ell,\, \ell_1, \, \ell_2 > 0$ with $\ell_1 + \ell_2 < \ell$. Then there exists a natural projection map
\[
\Lp_2([0,\ell],\Real)\otimes \Real^d \to (\Lp_2([0,\ell_1],\Real)\otimes \Real^d) \oplus (\Lp_2([\ell-\ell_2,\ell],\Real)\otimes \Real^d),
\]
which, by appropriate scaling, can be written as a map
\begin{equation}\label{projdef}
\pi_{\ell}^{\ell_1,\ell_2}: \Lp_2([0,1],\Real)\otimes \Real^d \to (\Lp_2([0,1],\Real)\otimes \Real^d) \oplus (\Lp_2([0,1],\Real)\otimes \Real^d),
\end{equation}
which is a bounded operator of norm $1.$\footnote{This map of course coincides with the projection of equation (\ref{eq:i6}) in the case when $\ell_1 + \ell_2 = \ell_3$, so we use the same notation.}

Since $\norm{\pi_{\ell}^{\ell_1,\ell_2}} =1,$ the operator $\pi_{\ell}^{\ell_1,\ell_2}$ induces an operator 
\begin{equation}\label{projdeffock}
\pi_{\ell}^{\ell_1,\ell_2} : \Fock \to \Fock \otimes \Fock
\end{equation}

\noindent which we continue to denote by $\pi_{\ell}^{\ell_1,\ell_2}.$

We now define mathematically the cut-off version of the function $I.$  Let $\epsilon >0$, $T>0$, $v \in (0,L_0/4),$ and define 
\[
\I_{M,\kappa}^{\epsilon, T,v} : C_c(\Real) \otimes_{\rm alg} \Lp_{2}([L_0,L_{\infty}]) \otimes_{\rm alg} \Fock \to \Cmpl
\] by 
\begin{multline}\label{defiold}
\I_{M,\kappa}^{\epsilon, T,v}(f \otimes g \otimes w) = 
 \negthickspace \int 
       \limits_{ \ell, \ell_1,\ell_2 \in [L_0,L_{\infty}]  }
            d\ell\, d\ell_1 \, d\ell_2 \,
                {g}_{\kappa}(\ell_1) \, {g}_{\kappa}(\ell_2)\,\overline{g}_{\kappa}(\ell) \,
                     2\delta_{1/v}(\ell_1 + \ell_2 - \ell)\theta(\ell-\ell_1 - \ell_2)\\
    \int_{-T}^{T} dt \, {f}^2(t) \overline{f}(t) 
        \bra 
                 e ^{ - \epsilon \Ham_{\ell,m}} 
          \pi_{\ell}^{\ell_1,\ell_2} p_M w,
             e ^{ - \epsilon \Ham_{\ell_1,m}}          p_M w
                \otimes      
             e ^{ - \epsilon \Ham_{\ell_2,m}}          p_M w    
        \ket_\Fock,     
\end{multline}
where $g_{\kappa} = g \star \delta _ {\kappa}$.

\begin{thm_}\label{thm:i2}
The function $\I_{M,\kappa}^{\epsilon, T,v}$ extends to a function $\I_{M,\kappa}^{\epsilon, T,v} \in \Lp_p(d\mu)$ for all $p \geq 2.$ The limit 
\[
\I^{\epsilon, T,v} := \lim _{M,\kappa \to \infty} \I_{M,\kappa}^{\epsilon, T,v}
\]
 exists in $\Lp_2(d\mu)$.
\end{thm_}

\begin{cor_}\label{cor:i3}
The function $\Z^{\epsilon,T,v}(\lambda) = \int_\Ban d\mu \, e ^ {i \lambda \Rep I^{\epsilon, T,v}}$ is a continuous function of $\lambda$ for all $\lambda \in \Real$.
\end{cor_}

\begin{rem}
In fact $\I^{\epsilon, T,v} \in \Lp_p(d\mu)$ for all $p \geq 2$, so that the function $\Z^{\epsilon,T,v}(\lambda)$ is smooth.
\end{rem}

\begin{rem}Note Corollary \ref{cor:i3} gives the existence of the ``Wick-rotated'' cut-off nonperturbative string partition function for pure imaginary values of the string coupling constant.  The parameters $\epsilon,T,L_0,L_{\infty}$ may be interpreted in terms of the Riemann surfaces appearing in the formal power series expansion as in (\ref{fpe}). In these terms the parameters require all ``tubes'' corresponding to edges of graphs to have length no less than $\epsilon$ and no greater than $T$, and width lying in the interval $[L_0, L_{\infty}]$. These Riemann surfaces are thus kept away from the boundary of the moduli space of curves, where Polyakov measure for the Bosonic string is known to diverge \cite{Wolpert}.\footnote{The light cone partition functions we have considered should be \cite{Dg} equal in the case $d=24$ to the Polyakov measure, so they should diverge as well.}
Unlike in the case of the superstring, these cutoffs cannot be removed in Bosonic string theory.  It is remarkable that these cutoffs, which the analysis requires in order to obtain a well-defined non-Gaussian integral, are precisely those that appear in the geometry of the Polyakov measure.  See Remark \ref{rem:3.3} for a discussion of the cutoff $v.$\end{rem}

\begin{rem}Note that finiteness of the limit $\lim \limits_{M,\kappa \to \infty} \norm{ \I^{\epsilon, T,v}_{M,\kappa}} ^2_2$, which corresponds to a sum of diagrams of genus two, implies existence of the partition function for all $\lambda \in \Real$. At least in this case, ``finiteness in genus two implies finiteness of the nonperturbative theory.''  In view of the recent work of d'Hoker and Phong \cite{DP} on finiteness of the superstring in genus two, this is very encouraging.\end{rem}

\subsection{Random Surfaces}\label{rsurf}

Formally the partition function $\Z^{\epsilon, T,v} (\lambda)$ may be expanded in a power series
\begin{equation}\label{eq:i.3.1}
\Z^{\epsilon, T, v}(\lambda) \sim \sum \limits_{n=0}^{\infty} \frac{(i \lambda)^{2n}}{(2n)!} \int_\Ban d\mu (\Rep \I ^{\epsilon, T, v})^{2n}
\end{equation}

Since $\I ^{\epsilon, T, v}$ is a cubic polynomial, each of the terms on the right side of equation \eqref{eq:i.3.1} may be written as a sum of terms, each of which corresponds to a directed trivalent ribbon graph $\Gamma$ with $2n$ vertices, each of whose edges $e$ is decorated with two real numbers $t_e \in [-T,T]$, $\ell_e \in [L_0,L_{\infty}]$. Let $G_{2n}$ denote the set of directed trivalent ribbon graphs with $2n$ vertices.
Given $\Gamma \in G_{2n}$ let $E(\Gamma)$ denote the set of edges of $\Gamma$. Thus 
\begin{multline*}
\negthickspace\negthickspace\negthickspace\int_\Ban d\mu \, (\Rep \I^{\epsilon, T, v})^{2n}\negthickspace = \negthickspace\sum 
                             \limits_{\Gamma \in G_{2n}}\negthickspace {|{\rm Aut}(\Gamma)|}
                             \int_{-T}^{T} \negthickspace\ldots \int_{-T}^{T} 
                               \prod \limits_{e \in E(\Gamma)} dt_e
                                                         \int_{L_0}^{L_{\infty}} \negthickspace\negthickspace \ldots  \int_{L_0}^{L_{\infty}} 
                               \negthickspace\prod \limits_{e \in E(\Gamma)} d\ell_e 
                               f_{\Gamma } (t_1, \ldots, t_{\abs{E(\Gamma)} }; \ell_1, \ldots, \ell_{\abs{E(\Gamma)} };v)
\end{multline*}

\noindent where $|{\rm Aut}(\Gamma)|$ is the order of the group of automorphisms of the directed ribbon graph $\Gamma$ and where the activities $f_\Gamma$ are given by the usual Feynman rules for Gaussian integrals.

Then we conjecture that the activities $f_{\Gamma}(t_1, \dots, t_{\abs{E(\Gamma)} }; \ell_1, \dots, \ell_{\abs{E(\Gamma)}};v)$ are related to Riemann surfaces, as follows.

Suppose we are given a directed trivalent ribbon graph $\Gamma$, each of whose edges $e\in E(\Gamma)$ is labeled by real numbers $t_e,\ell_e$. Each vertex of $\Gamma$ abuts three edges $e_i,e_j,e_k$ with 
\begin{equation}\label{eq:i.3.2}
\ell_i = \ell_j + \ell_k.
\end{equation}
\begin{figure}[htbp]
\begin{center}
\scalebox{.73}{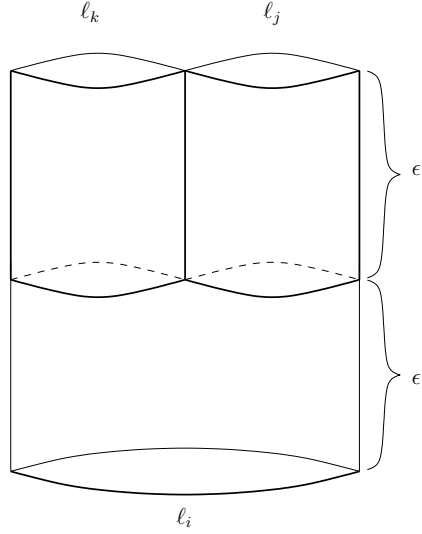}
\caption{A "plumbing  fixture" }\label{fig:1}
\end{center}
\end{figure}
We construct a Riemannian two-manifold with conical sigularities, which we denote by

$\Sigma (\Gamma, t_1, \ldots, t_{\abs{E(\Gamma)} }; \ell_1, \ldots, \ell_{\abs{E(\Gamma)} } ),$
by associating to each vertex of $\Gamma$ abutting edges $e_i,e_j,e_k$ a ``plumbing fixture'' consisting of three cylinders $C_1,C_2,C_3$ of widths $\ell_i, \ell_j, \ell_k$ with $\ell_i = \ell_j + \ell_k$ and each of length $\epsilon$, attached by gluing one of the boundary circles of each of $C_j$ and $C_k$ to one of the boundary circles of $C_i$ (see Figure \ref{fig:1}). We associate to each edge $e$ of $\Gamma$ a cylinder of length $t_e$ and width $\ell_e$, and we form a Riemann surface with conical singularities by gluing the cylinders corresponding to each edge to the ``plumbing fixtures'' of the vertices abutting $e.$  Given a connected trivalent ribbon graph $\Gamma$, with edges labeled as above, let $\det_{KK} (-\Delta_{\Sigma (\Gamma, t_1, \ldots, t_{\abs{E(\Gamma)} }; \ell_1, \ldots, \ell_{\abs{E(\Gamma)} } ) })$ denote the determinant of the Laplacian on $\Sigma (\Gamma, t_1, \ldots, t_{\abs{E(\Gamma)}}; \ell_1, \ldots, \ell_{\abs{E(\Gamma)}} )$ as defined by Kokotov and Korotkin \cite{kokkor}. Then
\begin{conj_}\label{conj4}
Let $\Gamma$ be a connected directed trivalent ribbon graph with $2n$ vertices. Suppose each edge $e$ of $\Gamma$ is labeled with real numbers $t_{e}, \ell_{e}$ satisfying the condition \eqref{eq:i.3.2}.  Then
 
\[
\lim \limits_{v \to 0} \lim \limits_{m \to 0}
     {v}{m^{d}}
     f_{\Gamma} (t_1, \ldots, t_{\abs{E(\Gamma)}}; \ell_1, \ldots, \ell_{\abs{E(\Gamma)}}; v)
    =  {\rm det}_{KK}^{-d/2} (-\Delta_{\Sigma (\Gamma, t_1, \ldots, t_{\abs{E(\Gamma)}}; \ell_1, \ldots, \ell_{\abs{E(\Gamma)}} ) } ).
\]
\end{conj_}

\begin{rem}Conjecture \ref{conj4} may be interpreted as saying that the Feynman diagrams of interacting string theory are partition functions of two dimensional quantum field theories on Riemann Surfaces.  Thus the partition function $Z^{\epsilon, T, v}$ would serve as a generating function for the partition functions of these quantum field theories (and, according to \cite{Dg}, for Polyakov measure on the moduli of curves in the case d=24) just as the finite-dimensional integrals of Kontsevich \cite{konts} serve as generating functions for intersection numbers on the moduli of curves.  We should add that the interacting string measure contains much more information than just the partition function; one should be able to find similar conjectures about correlations of vertex operators.  \end{rem}
\section{Free String Field Theory}\label{sec:free-sft}
In this section we prove Theorem \ref{thm:1} by constructing the free string measure $\mu$. 

For $\alpha \geq 0,$ let $\Fock^\alpha = e^{-\alpha \Ham_{L_0,m}} \Fock,$ equipped with the inner product 
\[
\bra \cdot ,\cdot \ket_{\alpha} = \bra e^{\alpha \Ham_{L_0,m}}\cdot, \,  e^{\alpha \Ham_{L_0,m}}\cdot \ket_\Fock
.\]

Consider the Hilbert space
\[
\Hilb^\alpha = \Hsp_{1}(\Real) \otimes \Lp_2([L_0,L_{\infty}]) \otimes \Fock^\alpha
\]
The quadratic form $C: \Hilb^\alpha \times \Hilb^\alpha \to {\mathbb C}$ given by
\[
\C(f'\otimes g' \otimes v', f \otimes g\otimes v) =  
\int_{L_0}^{L_\infty} d\ell g(\ell) \overline{g'}(\ell) \bra 
\left(\frac { - \frac{d^2}{dt^2} + \Ham_{\ell,m}^2 }{2\Ham_{\ell,m}}\right)^{1/2}
v' \otimes f'\, ,\left(\frac { - \frac{d^2}{dt^2} + \Ham_{\ell,m}^2 }{2\Ham_{\ell,m}}\right)^{1/2}\, v \otimes f \ket_{\Fock \otimes L_2(\Real)}
\]
gives rise to an inner product $\ip_C$ on $\Hilb^\alpha$ for all $\alpha > 0.$ Denote by $\Hilb$ the Hilbert space completion of $\Hilb^\alpha$ in the norm corresponding to  $\ip_C$; this is independent of $\alpha$ as long as $\alpha > 0.$

Let $\Hilb_M = \Hsp_{1}(\Real) \otimes \Lp_2([L_0,L_{\infty}]) \otimes \Fock_M$. Then the bounded operator $P_M : \Hilb_M \to \Hilb_M$ given by 
\[
P_M = \Bigl( \frac { - \frac{d^2}{dt^2} + \Ham_{\ell,m}^2 }{2\Ham_{\ell,m}} \Bigr)^{-1/2}  \circ \Bigl( - \frac{d^2}{dt^2} + 1 \Bigr)^{1/2}
\]
extends to an isometry $P: \Hilb_0 \to \Hilb.$

Let $\delta >0$. The operator $A=e^{-\delta \Ham_{L_0,m}} : \Fock \to \Fock$ is positive and trace class.  By Example \ref{ex:b2'}, the Sobolev norm $\norm{\;}_{-1}$ is a measurable norm on $\Lp_2([L_0,L_{\infty}]),$ given by the positive trace class operator $B = (-\Delta + 1)^{-1}$ on $\Lp_2([L_0,L_{\infty}]).$  It follows by Example \ref{prop:b4} that the norm $\norm{\;}_{0}$ given by  
\[
\norm{f}_{0}  =  {\rm sup}_{t\in \Real} \norm{f(t)}_{A \otimes B}
\]
is a measurable norm on $\Hilb_0.$  The completion of $\Hilb_0$ in this norm is
a subspace of $C(\Real; \Hsp_{-1}\left((L_0,L_\infty)\right) \otimes \Fock_A).$

Since $P : \Hilb_0 \to \Hilb$ is an isometry, the norm $\norm{\;}$ on $\Hilb$ given by
\[
\norm{x} = \norm{P^{-1} x}_0
\]
is a measurable norm on $\Hilb$. Let $\Ban$ denote the completion of $\Hilb$ in the norm $\norm{\;}.$  The projection $\pi_M: \Hilb \to \Hilb_M$ induces a projection on $\Ban,$ which we continue to denote by $\pi_M.$   Note that since $P_M$ is bounded, the norm $\norm{\;}_0$ is a measurable norm on $\Hilb_M,$ considered as a subspace of $\Hilb.$  Thus, although elements $\Psi \in \Ban$ are typically not continuous functions with values in $\Hsp_{-1}\left((L_0,L_\infty)\right) \otimes \Fock,$ the projections $\pi_M \Psi$ almost surely do lie in $C(\Real; \Hsp_{-1}\left((L_0,L_\infty)\right) \otimes \Fock).$

Gross' Theorem (Theorem \ref{thm:b1} of Appendix \ref{app:b}) then implies the following result, which is a slight restatement of Theorem 1:

\begin{thm}\label{thm:1}
Let $\Ban$ denote the completion of $\Hilb$ in the norm $\norm{\;}$. There exists a Gaussian Borel probability measure $\mu$ on $\Ban$ extending the natural cylinder set measure on $\Hilb$.

In particular, if $v,v' \in \Fock$ are eigenvectors of $\Ham_{\ell,m}$,  $g,g' \in \Cont^{\infty}([L_0,L_{\infty}])$, $t,t' \in \Real,$ the linear functional
\[
\Phi_{v,g,t} : \Hilb_M \to {\mathbb C}
\]
given by 

\[\Phi_{v,g,t}(f \otimes h \otimes w) = f(t) \bra h, g \ket_{L_2([L_0,L_\infty])} \bra w, v\ket_{\Fock} \]
extends to an element
\[
\Fi_{v,g,t} \in \Lp_2(d\mu);
\]
and
\begin{multline*}
\int_{\Ban} d\mu \, \overline{\Fi}_{v,g,t} \, \Fi_{v',g',t'} \\
  = \int_{L_0}^{L_{\infty}} d\ell \, \overline{g}(\ell) g'(\ell) 
      \bra
         \left(\frac {2 \Ham_{\ell,m}} { - \frac{d^2}{dx^2} + \Ham_{\ell,m}^2 }\right)^{1/2}v \otimes \delta(\cdot- t'),  
         \left(\frac {2 \Ham_{\ell,m}} { - \frac{d^2}{dx^2} + \Ham_{\ell,m}^2 }\right)^{1/2} v' \otimes \delta(\cdot - t)
      \ket_{\Fock \otimes L_2(\Real)}\\
   = \int_{L_0}^{L_{\infty}} d\ell \, \overline{g}(\ell) g'(\ell) 
      \bra
        v,
        e^{-\abs{t-t'} \Ham_{\ell,m}}v'
      \ket_\Fock.  
\end{multline*}
\end{thm}

\section{The interaction term}\label{sec:3}

In this section we define the interaction term $I^{\epsilon,T,v}_{M,\kappa}$ and prove the existence of the limit $\lim \limits_{M,\kappa \to \infty} I^{\epsilon,T,v}_{M,\kappa}$ in $\Lp_2(d\mu)$. The existence of the partition function $\Z^{\epsilon,T,v} (\lambda) = \int d\mu e ^{i \lambda \Rep \I^{\epsilon,T,v}}$ follows.

\subsection{The projection.}\label{subsec:3.1}

We begin with some technical results on the projection $ \pi_{\ell}^{\ell_1,\ell_2}.$

\begin{prop_section}\label{prop:3.1}
Let $\ell, \ell_1, \ell_2> 0$ with $ \ell_1 + \ell_2 < \ell$, $\alpha_1,\alpha_2, \alpha_3, m_1, m_2, m_3 > 0.$  Let $\pi_{\ell}^{\ell_1,\ell_2}: \Lp_2([0,1],\Real)\otimes \Real^d \to (\Lp_2([0,1],\Real)\otimes \Real^d) \oplus (\Lp_2([0,1],\Real)\otimes \Real^d)$
denote the projection (see equation (\ref{projdef})).  Then the family of operators
$$\Thet(\alpha_i, m_i, \ell_1, \ell_2 , \ell) \in\El\left(\Lp_{2}( [0,1],\Real )\otimes \Real^d,  (\Lp_{2}( [0,1],\Real )\otimes \Real^d)\oplus (\Lp_{2}( [0,1],\Real )\otimes \Real^d)\right)$$

\noindent given by
\[
\Thet(\alpha_i, m_i, \ell_1, \ell_2 , \ell) := 
                 \left(
                   e ^ {-\alpha_1 \sqrt{-(d^2/dx^2) + m_1^2} }
          \oplus   
                   e ^ {-\alpha_2 \sqrt{ -(d^2/dx^2) + m_2^2} } 
                \right) 
          \circ 
                   \pi_{\ell}^{\ell_1,\ell_2}
          \circ  
                   e ^ {-\alpha_3 \sqrt{ -(d^2/dx^2) + m_3^2} }
\]
is a continuously differentiable family of bounded operators.
\end{prop_section}

\begin{Proof} Differentiability in the $\alpha_i, m_i$ is clear; we compute the derivative with respect to $\ell_1$. We have
\[
\frac{1}{\epsilon} (   \pi_{\ell}^ {\ell_1  + \epsilon, \ell_2} f 
                        -  \pi_{\ell}^ {\ell_1            , \ell_2} f   )  (x)
 = \left(\frac{1}{\epsilon}
     \Bigl( 
        \sqrt{\frac{\ell_1 + \epsilon}{\ell}}f \bigl( \frac{\ell_1 + \epsilon}{\ell} x \bigr) - \sqrt{\frac{\ell_1}{\ell}}f \bigl( \frac{\ell_1}{\ell} x \bigr)
     \Bigr) , 0\right) .
\]

Thus

\begin{multline*}
\frac{1}{\epsilon}
   \Bigl( 
     \bigl(
       e ^ {-\alpha_1 \sqrt{ -(d^2/dx^2)+ m_1^2} }
          \oplus
       e ^ {-\alpha_2 \sqrt{ -(d^2/dx^2)+ m_2^2} }
     \bigr)
      (   \pi_{\ell}^ {\ell_1  + \epsilon, \ell_2} 
       -  \pi_{\ell}^ {\ell_1            , \ell_2}  )  f
   \Bigr) (y)\\
= 
\left(\frac{1}{\epsilon}
  \int_0^1 
   K_{\alpha_1,m_1}(y-x)   
   \Bigl( 
      \sqrt{\frac{\ell_1 + \epsilon}{\ell}} f \bigl( \frac{\ell_1 + \epsilon}{\ell} x \bigr) -\sqrt{\frac{\ell_1}{\ell}} f \bigl( \frac{\ell_1}{\ell} x \bigr)
   \Bigr) \, dx , 0\right) \\
= \left(\frac{1}{\epsilon}
 \left [
  \int_0^{\frac{\ell_1 + \epsilon}{\ell}}
   K_{\alpha_1,m_1}(y-\frac{\ell_1 + \epsilon}{\ell} \eta )   
   f(\eta) \sqrt{\frac{\ell}{\ell_1 + \epsilon} }
    \, d\eta
-
  \int_0^{\frac{\ell_1}{\ell}}
   K_{\alpha_1,m_1}(y-\frac{\ell_1}{\ell} \eta )   
   f(\eta) \sqrt{\frac{\ell}{\ell_1 } }
    \, d\eta
 \right ], 0\right),
\end{multline*}
where   $K_{\alpha,m}(x-y)$ is the smooth kernel of the operator $e ^ {-\alpha \sqrt{ -(d^2/dx^2) + m^2} }$ . 

Differentiability now follows from the smoothness of $K_{\alpha,m}$. Derivatives with respect to  $\ell_2$ and $\ell$ are similar.
\end{Proof}

\begin{cor_section}\label{cor:3.2}
Let $\ell, \ell_1, \ell_2 >0$ with $\ell_1 + \ell_2 < \ell$, let $m_i, \alpha_i > 0,$ and let  $\pi_{\ell}^{\ell_1,\ell_2} : \Fock \to \Fock \otimes \Fock$ denote the operator defined in equation (\ref{projdeffock}).  Then the family of operators given by 
\[
\hat{\Theta}(\alpha_i, m_i, \ell_1, \ell_2 , \ell) := 
              (e ^ {-\alpha_1 \Ham_{\ell_1,m_1} }
          \otimes    
              e ^ {-\alpha_2 \Ham_{\ell_2,m_2} } )
          \circ 
              \pi_{\ell}^{\ell_1,\ell_2}
          \circ e ^ {-\alpha_3 \Ham_{\ell_3,m_3} }  
\]
is a continuous family of operators in $\El(\Fock,\Fock \otimes \Fock).$
\end{cor_section}

\subsection{Definition of the interaction vertex and the proof of Theorem \ref{thm:i2}.}

To define the interaction vertex, we first cut off the string field $\Psi.$
Let $M > 0$. As we noted, elements of $\pi_M \Ban$ almost surely lie in $C(\Real;\Hsp_{-1}\left((L_0,L_\infty)\right) \otimes \Fock).$  We smooth in the second variable by the convolution 
$\star \delta_\kappa : C(\Real; \Hsp_{-1}\left((L_0,L_\infty)\right) \otimes \Fock) \to C(\Real; \Hsp_1\left((L_0,L_\infty)\right) \otimes \Fock)$ defined as follows.  For $g \otimes v \in \Hsp_{-1}\left((L_0,L_\infty)\right) \otimes \Fock)$, define $$(\delta_\kappa\otimes 1) \star (g \otimes v) = (\delta_\kappa \star g) \otimes v.$$  Now given $f \in C(\Real; \Hsp_{-1}\left((L_0,L_\infty)\right) \otimes \Fock),$ let
\[
(\star \delta_\kappa (f))(t) =  (\delta_\kappa\otimes 1) \star f(t) .\]

Then for $\Psi \in \Ban,$ and $M, \kappa > 0,$ define the cut-off string field $\Psi_{M,\kappa}$ by $$\Psi_{M, \kappa} = \star \delta_\kappa(\pi_M \Psi).$$

Let $\epsilon, T, M, \kappa > 0, v\in (0,L_0/4)$. The {\em cut-off interaction vertex} $I_{M,\kappa}^{\epsilon,T,v}$ is defined as follows.

{\noindent}   Given $\Ps_1,\Ps_2,\Ps_3 \in C(\Real; \Hsp_{1}\left((L_0,L_\infty)\right) \otimes \Fock),$ let
\begin{multline}\label{defj}
\J( \Ps_1,\Ps_2,\Ps_3)\ := \int_{-T}^{T} dt 
     \int_{L_0}^{L_{\infty}} d\ell_1
     \int_{L_0}^{L_{\infty}} d\ell_2
     \int_{L_0}^{L_{\infty}} d\ell \\
      2 \delta_{1/v}(\ell_1 + \ell_2 - \ell)  \theta(\ell- \ell_1 - \ell_2)
       \bra
e^ {- \epsilon H_{\ell,m}} 
         \pi_{\ell}^{\ell_1,\ell_2} 
	 \Ps_{1} ( \ell,t ) 
        ,
         e^{-\epsilon H_{\ell_1,m}}  \Ps_{2}(\ell_1,t ) \otimes  e^{-\epsilon H_{\ell_2,m}}  \Ps_{3}(\ell_2,t )
       \ket_\Fock.
\end{multline}
Let
\[\label{defi}
\I^{\epsilon,T,v}_{M,\kappa}(\Ps) = \J(\Ps_{M,\kappa},\Ps_{M,\kappa},\Ps_{M,\kappa}).
\]

It is clear that this function coincides with the function given by equation \eqref{defiold}.
Since $\I^{\epsilon,T,v}_{M,\kappa}$ is a polynomial cylinder function, $\I^{\epsilon,T,v}_{M,\kappa} \in \Lp_p(d\mu)$ for all $p \geq 1$. To prove Theorem \ref{thm:i2}, we must show that
\[
\lim_{M,M',\kappa,\kappa'\to \infty} \norm{\I^{\epsilon,T,v}_{M,\kappa} - \I^{\epsilon,T,v}_{M',\kappa'}}^2_2 = 0.
\]

Now 
\begin{equation}\label{eq:3.1}
\norm{\I^{\epsilon,T,v}_{M,\kappa} - \I^{\epsilon,T,v}_{M',\kappa'}}^2_2 
= 
 \int_\Ban d\mu \left( \abs{ \I^{\epsilon,T,v}_{M,\kappa}}^2 + \abs{ \I^{\epsilon,T,v}_{M',\kappa'}}^2
-
 2\Rep 
  \left( (\I^{\epsilon,T,v}_{M,\kappa})^*
   \I^{\epsilon,T,v}_{M',\kappa'}\right)\right).
  \end{equation}
Let $\delta \Ps = \Ps_{M',\kappa'} - \Ps_{M,\kappa}$. Then by rearranging terms in the expression on the right hand side of \eqref{eq:3.1} we have:
\begin{multline}\label{eq:3.2}
 \norm{   \I^{\epsilon,T,v}_{M,\kappa}   -        \I^{\epsilon,T,v}_{M',\kappa'}  }^2_2
= 
 \int_\Ban d\mu (\Ps) \,
       \left[
          \overline{\J}(\Ps_{M',\kappa'},\Ps_{M',\kappa'},\Ps_{M',\kappa'})-            \overline{\J}(\Ps_{M,\kappa},\Ps_{M,\kappa},\Ps_{M,\kappa}) \right]   \\
         \left[ 
              \J(\delta \Ps      ,\Ps_{M',\kappa'},\Ps_{M',\kappa'}) 
            + \J(\Ps_{M,\kappa}  ,\delta \Ps      ,\Ps_{M',\kappa'})
            + \J(\Ps_{M,\kappa}  ,\Ps_{M,\kappa}  ,\delta \Ps) 
         \right]
\end{multline}

Consider a typical term in \eqref{eq:3.2}, given by
\[
\int_\Ban d\mu (\Ps)   \overline{\J}(\Ps_{M',\kappa'},\Ps_{M',\kappa'},\Ps_{M',\kappa'} )
               \, \J (\delta \Ps      ,\Ps_{M,\kappa}  ,\Ps_{M,\kappa}   ).
\]

If $\kappa, \kappa'$ are sufficiently large, the condition $v < L_0/4$ guarantees that this integral is given by a sum of terms corresponding to two ribbon graphs.  We have
\begin{multline}\label{vani}
\int_\Ban d\mu (\Ps)   \overline{\J}(\Ps_{M',\kappa'},\Ps_{M',\kappa'},\Ps_{M',\kappa'} )
               \, \J (\delta \Ps      ,\Ps_{M,\kappa}  ,\Ps_{M,\kappa}   )=\\
\int_{-T}^{T} \int_{-T}^{T} dt dt' \int_{L_0}^{L_{\infty}} \ldots \int_{L_0}^{L_{\infty}}
      d\ell_1 \,d\ell_2 \,d\ell \; d\ell'_1 \,d\ell'_2 \,d\ell' \;  d\lambda_1 \,d\lambda_2 \,d\lambda\\
4 \theta (\ell-\ell_1 - \ell_2) \, \theta (\ell'-\ell'_1 - \ell'_2)
 \delta_{1/v} (\ell_1 + \ell_2 - \ell) \, \delta_{1/v} (\ell'_1 + \ell'_2 - \ell')
     \delta_{\kappa'}(\lambda - \ell') \,  \delta_{\kappa'}(\lambda_1 - \ell_1)\,  \delta_{\kappa'}(\lambda_2 - \ell_2)\,\\
 \Bigl \lbrace
 \left (     \delta_{\kappa'}(\lambda - \ell) - \delta_{\kappa}(\lambda - \ell) 
  \right ) 
\trace
\Bigl [ 
 \left(\pi_{M'} \pi_M \right) \, e ^{- \left( \abs{t -t'}\Ham_{\lambda,m} +  \epsilon  \Ham_{\ell,m} + \epsilon \Ham_{\ell',m}\right)} \, \left(\pi_{\ell} ^ {\ell_1,\ell_2}\right)^*\\
                    \Bigl (
 \delta_{\kappa'}(\lambda_1 - \ell'_1)\,\delta_{\kappa'}(\lambda_2 - \ell'_2) 
e ^{-\left( \abs{t -t'}\Ham_{\lambda_1,m}+  \epsilon \Ham_{\ell_1,m} +\epsilon \Ham_{\ell'_1,m}\right)}  
                    \otimes 
                         e ^{- \left( \abs{t -t'}\Ham_{\lambda_2,m} + \epsilon \Ham_{\ell_2,m} +\epsilon \Ham_{\ell'_2,m}\right)}+\\
\delta_{\kappa'}(\lambda_1 - \ell'_2)\,\delta_{\kappa'}(\lambda_2 - \ell'_1) E 
\left(
 e ^{-\left( \abs{t -t'}\Ham_{\lambda_1,m}+  \epsilon \Ham_{\ell_1,m} +\epsilon \Ham_{\ell'_2,m}\right)}  
                    \otimes 
                         e ^{- \left( \abs{t -t'}\Ham_{\lambda_2,m} + \epsilon \Ham_{\ell_2,m} +\epsilon \Ham_{\ell'_1,m}\right)} 
\right)
\Bigr )
  \left( \pi_{M'}\otimes \pi_{M'} 
             \right ) 
 \left( \pi_{\ell'}^{\ell'_1,\ell'_2} 
             \right) 
\Bigr ] \\                                                                                                 
+
  \delta_{\kappa'}(\lambda - \ell) \, 
\trace
\Bigl [ 
 \pi_{M'}\left(\pi_{M'} - \pi_{M} \right)
\, e ^{- \left( \abs{t -t'}\Ham_{\lambda,m} +  \epsilon  \Ham_{\ell,m} + \epsilon \Ham_{\ell',m}\right)} \, \left(\pi_{\ell} ^ {\ell_1,\ell_2}\right)^*\\
 \Bigl (
 \delta_{\kappa'}(\lambda_1 - \ell'_1)\,\delta_{\kappa'}(\lambda_2 - \ell'_2) 
e ^{-\left( \abs{t -t'}\Ham_{\lambda_1,m}+  \epsilon \Ham_{\ell_1,m} +\epsilon \Ham_{\ell'_1,m}\right)}  
                    \otimes 
                         e ^{- \left( \abs{t -t'}\Ham_{\lambda_2,m} + \epsilon \Ham_{\ell_2,m} +\epsilon \Ham_{\ell'_2,m}\right)}+\\\delta_{\kappa'}(\lambda_1 - \ell'_2)\,\delta_{\kappa'}(\lambda_2 - \ell'_1)E 
\left(
 e ^{-\left( \abs{t -t'}\Ham_{\lambda_1,m}+  \epsilon \Ham_{\ell_1,m} +\epsilon \Ham_{\ell'_2,m}\right)}  
                    \otimes 
                         e ^{- \left( \abs{t -t'}\Ham_{\lambda_2,m} + \epsilon \Ham_{\ell_2,m} +\epsilon \Ham_{\ell'_1,m}\right)} 
\right)
\Bigr )
 \left( \pi_{M'}\otimes \pi_{M'} 
             \right ) 
 \left( \pi_{\ell'}^{\ell'_1,\ell'_2} 
             \right) 
\Bigr ] \Bigr \rbrace
\end{multline}

\noindent where $E:\Fock \otimes \Fock \to \Fock \otimes \Fock$ is the exchange map $E(a\otimes b) = b\otimes a.$

The vanishing of this expression in the limit $\kappa,\kappa',M,M' \to \infty$ follows by the trace class property of the heat kernel $e^{-t \Ham_{\ell,m}}$ and the differentiability of Corollary \ref{cor:3.2}. The other terms in equation \eqref{eq:3.2} are similar.
\begin{rem}\label{rem:3.3}

Note that it is not possible to take the limits $\epsilon,T \to \infty, v\to 0$ in the same way. The cutoffs $\epsilon,T$ would appear to be necessary since in the formal power series expansion of $\Z^{\epsilon,T,v}(\lambda) = \int_\Ban d\mu \exp( i \lambda \Rep \I^{\epsilon,T,v})$ they prevent the Riemann surfaces that appear from approaching the boundaries of the moduli space of Riemann surfaces. The role of the cut-off $v$ is more obscure. In the formal power series expansion of $\Z^{\epsilon,T,v}(\lambda),$ each activity $f_\Gamma$ grows as
$$f_\Gamma(\dots;v) \sim v^{-|H_0(\Gamma)|}$$

\noindent as $v \to 0$ (This can be seen explicitly in the integral (\ref{vani}) which corresponds to the sum of two connected graphs). Standard considerations lead us to expect that $\log \Z^{\epsilon, T, v}(\lambda) \sim v^{-1} F^{\epsilon,T}(\lambda),$ where $F^{\epsilon,T}(\lambda)$ is an analog of the free energy.  The behavior of this cutoff is therefore similar to that of the finite volume cutoff in quantum field theory, with $v^{-1}$ playing the role of the volume.  In the case of quantum field theory, without such a finite volume cutoff, interacting field measures are not perturbations of Gaussian measures.  In that case,  the finite volume cutoff can often be removed by methods of Statistical Mechanics.  I do not know if such methods will work for string field theory.

Note also that the cutoff $v> 0$ is required even in the field theory limit (which appears if the Fock space cutoff $M$ is sufficiently small) due to the fact that there are no derivatives with respect to $\ell$ appearing in the action.  In mathematical language, the free string measure becomes the tensor product of a massive analog of Wiener measure with a white noise measure, and so is supported on distributions.  Thus the cubic interaction must be regularized even in this limit.  
\end{rem}
\section{Concluding Remarks}
\subsection{Relation to standard ideas in string theory}
It is important to note the differences in principle between the nonperturbative partition function $Z^{\epsilon, T, v} (\lambda)$ and ideas arising in physics.

\begin{itemize}
\item We work in imaginary time;
\item The string coupling constant is pure imaginary;
\item We placed an infrared cutoff $m>0$ on the propagators;
\item We impose a moduli space cutoff given by parameters $\epsilon, T, L_0, L_\infty$;
\item Our action is second order in derivatives in $t,$ unlike the action of \cite{KK} which is first order;
\item We require an additional smoothing of the interaction, given by taking a nonzero value of the parameter $v.$
\end{itemize}

Despite these differences we hope our construction gives some insight into the mathematics of string theory.

\subsection{The field theory limit and promotion}\label{promotion}

Let $\pi_\Omega: \Fock \to \Fock$ denote the projection onto the vacuum $\Omega \in \Fock.$  This map induces a projection on $\Ban$ which we also denote by
$\pi_\Omega.$  The string field theory we have constructed has a quantum field theory limit obtained by replacing the interaction term $I^{\epsilon, T, v}(\Psi)$ by the projected interaction $I^{\epsilon, T, v}(\pi_\Omega\Psi).$  In this limit our string field theory is essentially a quantum field theory in $d+1$ dimensions.

On the other hand, one can imagine ``promoting'' the parameter $t$ to a field---that is, to a distribution in ${\mathcal S}'(S^1)$---and replacing the operator $(-d^2/dt^2 + 1)$ with the quantum mechanical Hamiltonian $H_{\ell, m}.$  It would be interesting to construct the corresponding measure.  I do not know if this construction would shed any light on the relation between superstrings and M theory.

\subsection{Twists and the moduli space of curves}\label{twists}
In Section \ref{rsurf} we conjectured that the activities $f_\Gamma(t_1,\dots, \ell_1,\dots;v)$ are related to determinants of Laplacians on Riemann surfaces.  It is expected \cite{Dg} that those determinants should in turn, in the case when $d=24,$ be related to Polyakov measure.  With this in mind it is natural to ask whether the surfaces $\Sigma(\Gamma,t_1,\dots,\ell_1,\dots)$ form a cover of the moduli of curves of genus $n+1$ as $\Gamma$ varies over $G_{2n}$ and the parameters $t_e, \ell_e$ vary over $[0,\infty)$ (here we are taking the limit as $\epsilon \to 0$ and $T \to \infty$).  

A quick dimension count shows that this cannot be the case; the surfaces formed by our procedure are parametrized by $3n$ parameters, while the dimension of the moduli space is $6n.$  A variant of our construction is the following.  For $\theta \in [0,2\pi]$, let $R(\theta): S^1 \to S^1$ denote the rotation.  This map induces a commuting family of unitary operators on $L_2(S^1)$, and hence a commuting family $R_\Fock(\theta)$ of unitary operators on $\Fock.$  Since $\Ps_{M,\kappa}$ is a function with values in $\Fock,$ we may define

\[\label{defrota}
\hat{I}^{\epsilon,T,v}_{M,\kappa}(\Ps) = \int_0^{2\pi} \int_0^{2\pi} \int_0^{2\pi} d\theta_1 d\theta_2 d\theta_3 \J(R_\Fock({\theta_1})\Ps_{M,\kappa},R_\Fock({\theta_2})\Ps_{M,\kappa},R_\Fock({\theta_3})\Ps_{M,\kappa}).
\]

The existence of the limit of $\hat{I}^{\epsilon,T,v}_{M,\kappa}(\Ps)$ as $M, \kappa \to \infty$ follows by the methods used to prove Theorem \ref{thm:i2}.  The analog of Conjecture \ref{conj4} now includes Riemann surfaces formed by attaching tubes to plumbing fixtures with twists between $0$ and $2\pi,$ and those twists give an additional $3n$ parameters.  In view of the work of Giddings and Wolpert \cite{gw}, it is possible that these singular surfaces provide a cover of the moduli space of curves.

\subsection{Open Strings, unoriented strings, and analytic semigroups}

It should be possible to repeat our construction in the case of open strings (with Neumann boundary conditions).  Such a construction should be related to moduli spaces of Riemann surfaces with boundaries, and a conjecture similar to Conjecture \ref{conj4} should exist, involving an appropriate determinant of the Laplacian on manifolds with boundary, with Neumann boundary conditions.

Likewise, replacing the complex Banach space $\Ban$ with a real Banach space should result in unoriented surfaces arising in the formal power series expansion.

It would also be interesting to investigate whether our measures, in the free case or the interacting case in the limit $T\to \infty,$ correspond to a reasonable semigroup acting on a Hilbert space.  In the free case ideas of this type have been studied in a different context by Dimock \cite{dimock}.

\appendix
\section{Fock space.}\label{app:a} 
We summarize here some basic information about Fock Space and path integrals.  See \cite{GJ} for more information and proofs.

Let $\Hilb = \Lp_2(\Str,\Real) \otimes \Real^d$. The symmetric tensor algebra $\Sym^{*} \Hilb$ has a natural inner product normalized so that 
\[
\norm{f \otimes \ldots \otimes f} = \norm{f}^n , \quad f\in \Hilb.
\]
The Fock space $\Fock$ is the completion of $\Sym^{*} \Hilb\otimes {\mathbb C}$ in this norm. The vacuum $\Omega$ is given by $\mathbf{1} \in \Sym^{*}\Hilb$. Let $e_1,\dots, e_d$ be a basis for $\Real^d.$ For each $p \in 2\pi \Intg, \, i = 1, \ldots ,d$, we define an annihilation operator $a_i(p)$ on a dense subset of $\Fock$ by 

$$a_i(p) (f_1 \otimes \dots \otimes f_n) = \sum_{j=1}^n 
\bra f_j, \exp(2 \pi i  p~ \cdot) \otimes e_i \ket f_1 \otimes \dots \otimes \hat{f_j} \otimes \dots \otimes f_n$$ Then
\[
 \left[ a_i(p), a_j(q) \right] =  \left[ a^{*}_i(p), a^{*}_j(q) \right]=0 
\quad 
  \text{and} 
\quad 
 \left[ a_i(p), a^{*}_j(q) \right] = \delta_{i,j} \delta_{p,q},
\]
where  $\delta_{a,b}$ is the Kronecker delta.

In terms of these operators, the Hamiltonian $\Ham_{\ell,m}$ (where $\ell, m > 0$) is given by 
\[
\Ham_{\ell,m} = \sum \limits_{i = 1}^{d} 
                \sum \limits_{k \in 2 \pi \Intg}
                     \left( \sqrt{ \left(\frac{k}{\ell}\right)^2 + m^2}\right) \, a_i^*(k) a_i(k).
\]

The operators $\Ham_{\ell,m}$ are unbounded self-adjoint positive operators on $\Fock$, with compact resolvent. Furthermore, for $t > 0$ the operator $e^{-t \Ham_{\ell,m}}$ is a trace class operator on $\Fock$, and
\[
e^{-t \Ham_{\ell^{'},m}} \geq e^{-t \Ham_{\ell,m}},
\]
whenever $\ell^{' }> \ell.$

An alternative description of $\Fock$ is given by the Schrodinger (or loop space) representation:  
Let $d\mu_{\ell,m}$ denote Gaussian measure on $\Sch ^{'}(\Str_{\ell},\Real)\otimes \Real^d$ with covariance 
\[
\left( -\frac{d^2}{dx^2} + m^2 \right)^{-1/2}.
\]
Then (see \cite{GJ}), $\Fock \simeq \Lp_2(\Sch^{'}(\Str_{\ell},\Real)\otimes\Real^d, d\mu_{\ell,m})$.

The relation to two dimensions is given by the Feynman-Kac formula, of which the simplest case is the following: Let $d\nu_{\ell,C}$ denote Gaussian measure of covariance $C = (-\Delta + m^2)^{-1}$ on $\Sch^{'}(\Str_{\ell}\times \Real,\Real)\otimes \Real^d$.

Then if $f,f' \in \Cont^{\infty}(\Str_{\ell},\Real)\otimes\Real^d$, $t,t' \in \Real$, the function 
$\Fi_{f,t}: \Sch'(\Str_{\ell} \times \Real;\Real^d) \to {\Real}$,
given by 
\[
\Fi_{f,t}(\psi) = \psi( \cdot, t)(f)
\]
extends to an element of $\Lp_{p}(d\nu_{\ell,C})$ for all $p \geq 1$, and 
\[
\int_{\Sch'(\Str_{\ell} \times \Real;\Real^d)} d\nu_{\ell,C}(\psi) \,{\Fi}_{f,t}(\psi) \Fi_{f',t'} (\psi)
         = \frac{1}{2}\bra 
             f ,
             \frac{e ^{ -\abs{t -t'} \sqrt{ -(d^2/dx^2) + m^2}}}{ \sqrt{ -(d^2/dx^2) + m^2} } f'
           \ket_{L_2(S^1_\ell \times \Real;\Real)\otimes \Real^d}  .
\]
\section{Abstract Wiener Spaces}\label{app:b}

Let $\Hilb$ be a separable Hilbert space. If $V \subset \Hilb$ is a finite-dimensional subspace of $\Hilb$ and $\pi_V : \Hilb \to V$ is the projection, a {\em cylinder set based on $V$} is a set of the form $\pi^{-1}_V(U) $ where $U \subset V$ is a Borel set. Similarly, a {\em cylinder function} is a function of the form $\pi_V^*f$ where $f: V \to \Cmpl$ is a Borel measurable function. Since every finite dimensional Hilbert space $V$ is isometric to $\Cmpl^n$ for some $n$, each such space is equipped with a natural Gaussian probability measure $\mu_{V}$. Thus the Hilbert space $\Hilb$ is equipped with a measure $\mu_{\Hilb}$ on cylinder sets given by 
\[
\mu_{\Hilb} ( \pi_{V}^{-1}(U)) = \mu_V (U).
\]
However, this measure does not extend, in the case of infinite-dimensional Hilbert spaces, to a countably additive measure on the Borel sets of $\Hilb$. Instead, we have the following construction due to Gross \cite{Gross}.

A norm $\norm{\;}_1$ on a Hilbert space is called {\em measurable} if for every $\epsilon > 0$, there exists a finite-dimensional space $V_{\epsilon} \subset \Hilb$ such that whenever $W \subset \Hilb$ is a finite dimensional space orthogonal to $V_{\epsilon},$
\[
\mu_{W}\left( \{ x \in W: \norm{x}_1 > \epsilon \} \right) < \epsilon.
\]

Let $\Ban$ denote the completion of $\Hilb$ in the norm $\norm{\,}_{1}$. Then Gross' theorem is
\begin{thm_section}\label{thm:b1}
The cylinder set measure $\mu_{\Hilb}$ extends to a Borel measure $\mu$ on $\Ban$. The measure $\mu$ is characterized by the following property. Any element $\Ps \in \Ban ^* \subset \Hilb$, may be considered as a function $\Fi_{\Ps}$ on $\Ban$. Then $\Fi_{\Ps} \in \Lp_2(d\mu)$ and for $\Ps, \Ps' \in \Ban^* \subset \Hilb$
\[
\int_\Ban d\mu \, \overline{\Fi}_{\Ps} \Fi_{\Ps'} = \bra \Ps, \Ps' \ket_\Hilb.
\]
\end{thm_section}
\begin{ex}\label{ex:b2}[Gaussian measure in one dimension.]
Let $\Hilb = \Hsp_{1}(\Real)$, and let $\norm{\;}_\infty$ denote the uniform norm (recall that by the Sobolev embedding theorem elements of $\Hsp_1(\Real)$ are bounded continuous functions).  Then $\norm{\;}_\infty$  is a measurable norm on $\Hilb,$ and the resulting Banach space is $\Ban = (C_b(\Real), \norm{\;}_\infty).$  This gives a massive analog of Wiener measure.
\end{ex}
\begin{ex}\label{ex:b2'}[White noise measure in one dimension]
Let $\Hilb = L_2([a,b])$, and let $\norm{\;}_{-1}$ be the Sobolev $(-1)$-norm.  Then $\norm{\;}_{-1}$ is measurable, and $\Ban = \Hsp_{-1}\left((a,b)\right)$. The resulting measure is white noise measure on the interval $[a,b].$
\end{ex}
\begin{ex}\label{ex:b3}
Let $\Hilb$ be a separable Hilbert space and let $A$ be a positive trace class operator on $\Hilb$. Then the norm given by $\norm{x}_A = \bra Ax,Ax \ket_\Hilb ^{1/2}$ is a measurable norm on $\Hilb$.  Denote the completion of $\Hilb$ in this norm by $\Hilb_A.$
\end{ex}

\begin{ex}\label{prop:b4}
Let $\tilde{\Hilb}$ be a separable Hilbert space, and let $\Hilb=\Hsp_{1}(\Real)\otimes\tilde{\Hilb}.$   This is a space of $\Hsp_{1}$-functions on $\Real$ with values in $\tilde{\Hilb}.$  Let $A$ be a positive trace-class operator on $\tilde{\Hilb}.$  For $f \in \Hilb,$ define 
$\norm{f} = {\rm sup}_{t \in \Real} \norm{f(t)}_A.$ Then $\norm{\;}$ is a measurable norm on $\Hilb,$ and the completion of $\Hilb$ is a subspace of the space $C(\Real;\tilde{\Hilb}_A)$ of continuous functions $f:\Real \to \tilde{\Hilb}_A$ with ${\rm sup}_{t\in \Real} \norm{f(t)}_A < \infty.$
\end{ex}

\end{document}